\documentclass{aa} %standard style
\input epsf

\usepackage{graphicx}
\usepackage[mathcal]{eucal}
\usepackage{amssymb}
\usepackage{amsmath}
\usepackage{natbib}
\usepackage[T1]{fontenc}
\def\ds {$\delta$\,Scuti}
\def\gd {$\gamma$\,Doradus}

\def\teff {{T_{\mathrm{eff}}}}

\def\cd {d$^{-1}$}

\begin{document}

%\topmargin -2.0truecm
%\topmargin -2.5truecm

\title{Strong interactions between $g$- and $p$-modes in the hybrid \gd-\ds\ CoRoT star ID\,105733033
\thanks{The CoRoT space mission was developed and is operated by the French space agency CNES, 
with participation of ESA's RSSD and Science Programmes, Austria, Belgium, Brazil, Germany, and Spain.}}

\author{E. Chapellier \inst{1}, P. Mathias \inst{2,} \inst{3}, W.W. Weiss \inst{4}, D. Le Contel \inst{1},
J. Debosscher \inst{5}
}  

\institute{
Laboratoire Lagrange, UMR7293, Universit\'e de Nice Sophia-Antipolis, CNRS, 
Observatoire de la C\^ote d'Azur, 06300 Nice, France
\and
Universit\'e de Toulouse; UPS-OMP; IRAP;  F-65000 Tarbes, France
\and
CNRS; IRAP; 57, Avenue d'Azereix, BP 826, F-65008 Tarbes, France
\and
University of Vienna, Institute for Astronomy, T\"urkenschanzstrasse 17, A-1180 Vienna, Austria
\and
Instituut voor Sterrenkunde, KU Leuven, Celestijnenlaan 200D, B-3001 Leuven, Belgium 
} 

%\offprints{E. Chapellier}

\date{Received date; accepted date}

\authorrunning{E. Chapellier et al.}
\titlerunning{interactions between $g$- and $p$-modes}

% if using the new \documentclass{aa} o similar

% \abstract{}{}{}{}{} 
% 5 {} token are mandatory
 
\abstract
{The presence of stellar $p$- and $g$-modes allows us to test stellar 
structure models in great detail from the core to the envelope. 
As the driving mechanisms are not yet fully understood, the first important step is to provide 
clear evidence of these pulsation modes. 
}
{Recent space missions have confirmed that the \gd\ and the \ds\ observational instability strips 
overlap and consequently that many stars may be hybrids. 
CoRoT\,ID\,105733033 is an excellent example of these hybrid pulsators as it shows $g$- and 
$p$-modes with almost similar amplitudes in two clearly distinct frequency domains.
We present a detailed frequency analysis of the CoRoT star ID\,105733033, which is obtained with a 
classical Fourier analysis.
} 
{After removing residual instrumental effects from the CoRoT light curve of N2 level, frequencies 
with an amplitude as small as 0.1\,mmag were determined with Period04 and SigSpec up to 50\,\cd, 
although if deemed necessary lower amplitudes and higher frequencies were also investigated. 
The frequency spectrum of CoRoT\,ID\,105733033 clearly consists of two distinct ranges, 
which are typical of \gd\ and \ds\ pulsation. 
Focus was placed on the identification of linear combinations and frequencies due to the coupling 
between \gd\ and \ds\ modes. 
}
{We detect 198 \gd\ type frequencies in the range $[0.25;4]$\,\cd, of which 180 are not 
combination frequencies, and 24 of them are separated by a constant period-interval 
$\Delta P=0.03074$\,d. 
According to the asymptotic theory, these 24 frequencies correspond to a series of $g$-modes 
of the same $\ell$-degree and different radial orders $n$.
We also detect 246 \ds\ type frequencies in the range $[10.1;63.4]$\,\cd. 
The dominant frequency $F=12.6759$\,\cd\ was identified as the fundamental radial mode. 
Our most noteworthy result is that all the main \gd\ frequencies $f_i$ are also detected in the 
\ds\ domain as $F \pm f_i$ with four times smaller amplitudes. 
Once these frequencies were removed, only 59 can be considered as individual \ds\ frequencies.
}
% conclusions heading (optional)  
{A coupling between $g$- and $p$-modes is proposed to be a tool for detecting $g$-modes 
in the Sun, but this coupling has never yet been observed. 
Our present study may be valuable input to theoretical studies, addressing the mutual 
influence of $g$- and $p$-mode cavities and the deviation from classical theory. 
Furthermore, we identify a sequence of $g$-modes belonging to the same $\ell$ but with consecutive 
orders $n$.
}


\maketitle

\section{Introduction}

The knowledge of solar and stellar interiors has improved tremendously in the past few decades 
due to both theoretical work such as new opacity tables \citep{cmr92} for the destabilisation of 
$\beta$\,Cephei and SPB stars and the convection/pulsation coupling for \gd\ stars \citep{gkb00,dgg04}, 
and new observing facilities such as the space missions CoRoT \citep{betal06}, MOST \citep{wmk03}, 
and Kepler \citep{bkd97,cab08}.

Until recently, \ds\ and \gd\ stars seemed to be clearly distinct, even if their respective instability 
strips overlap within a small region.
A few hybrid star candidates have been identified e.g. by \citet{hf05}, \citet{rmc06}, and \citet{h09}. 
The hybrid stars have become extremely interesting targets because $p$- and $g$-modes probe different 
regions of the stellar interiors. 
The presence of both types of modes in a same star provides complementary model constraints.

With the availability of space observations, considerably lower-amplitude pulsation modes have become 
detectable compared with ground observations. 
The instability strips (IS) of \ds\ and \gd\ stars have been found to overlap and their respective hot and cool 
borders have become seemingly less well-defined. 
In addition, many CoRoT and Kepler stars do not show in their pulsation spectra a clear distinction between their
\gd\- and \ds-frequency regions. 

In this paper, we discuss the very interesting case of a star being unambiguously hybrid, with a 
frequency gap between 4\,\cd\ and 10\,\cd, and indications of strong interactions between the $p$- and 
$g$-modes. 
The data are described in Sect.\,2, and the frequency analysis is presented in Sect.\,3. 
The frequency spectrum is discussed in Sect.\,4 for the low-frequency region (\gd\ domain) and in Sect.\,5 
for the high-frequency region (\ds\ domain). 
A theoretical interpretation of the interaction of the \gd-\ds\ modes is given in Sect.\,5. 
The effects of the rotation are discussed in Sect.\,6 and some conclusions are provided in Sect.\,7.

\section{The CoRoT data}

The observations of CoRoT\,ID\,105733033 were collected during CoRoT's second long run, LRc02, 
which targeted the Galactic center.
We used the reduced N2 light curves \citep{abb09} throughout this paper. 
The observations lasted 145 days, from 2008, April 15$^{\rm th}$ to September 7$^{\rm th}$. 
Among the 388950 measurements obtained with a temporal resolution of 32\,s, we retained only the 
345908 points flagged `0' by the CoRoT pipeline, the other measurements being affected by instrumental 
effects such as straylight, cosmic rays, and perturbation by Earth eclipses.

The light curve is recorded in "chromatic" light i.e., in three non-calibrated bands unevenly spanning 
the wavelength interval 370-950\,nm, obtained by the insertion of a prism just in front of the 
exo-planet CCD. 
To increase the signal-to-noise ratio (S/N), we added the three components of the colour bands to a "white" 
light curve. 
Unless specified otherwise, we refer in the following to the "white" light curve.

\begin{figure}
\centering
\includegraphics[width=8cm]{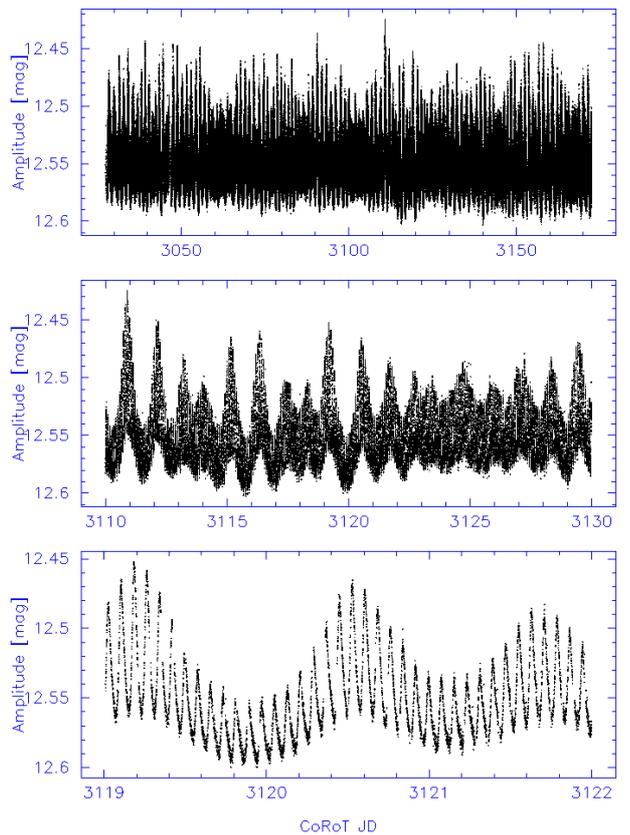}
\caption[]{Light curve of the star CoRoT\,ID\,105733033 corrected for long-term trends, jumps, and outliers 
(see text) with different timescales. 
From top to bottom, the complete light curve over 145\,d, then a subset over 22\,d and finally a zoom into a 
3.5\,d subset.}
\label{fig01}
\end{figure}

CoRoT data are known to be affected by several instrumental effects, such as long-term trends and jumps 
due to cosmic rays \citep{abb09}. 
In addition, many individual measurements can be considered as outliers. 
The data of the most significant outliers (mainly high-flux data points caused by cosmic ray impacts) 
were removed by an iterative procedure during the Fourier analysis.
Jumps were corrected by applying simple vertical shifts to the data. 
The resulting light curve is represented at different timescales in Fig.\,\ref{fig01} and clearly contains
both low and high frequency components.
As noted by \citet{b11} for other high amplitude \gd\ stars, the light curve displays sometimes a Blazhko-like 
behaviour: the maxima vary far more than the minima. 
This effect can also be observed for our star. 
In the following, the timescale is labeled in units of the CoRoT Julian day (JD), 
where the starting CoRoT JD corresponds to HJD\,2445545.0 (2000, January 1$^{\rm st}$ at UT 12:00:00).
%figure 1

Information provided by the EXODAT database \citep{dmm09} indicates that the star has an A5V spectral type 
and magnitude $V=12.8$. 
The conversion of the flux values to magnitudes of CoRoT\,ID\,105733033 is calibrated to a CoRoT magnitude 
of $C=12.543$. 
The BEST survey of the LRc02 field classifies the star as a pulsating star of SX\,Phoenicis type \citep{kfr09} 
and 2MASS provides (J=11.445, H=11.205, K=11.094). 
The CoRoT contamination factor is very small (1.7\,\%), hence we assume that the signal originates entirely from 
CoRoT\,ID\,105733033.

\section{Frequency analysis of the CoRoT dataset}

The frequency analysis was performed on the filtered data using two independent
methods: Period04 \citep{lb05} and SigSpec \citep{r07}.
We first used the Period04 package and searched the interval $[0;100]$\,\cd\ i.e., well 
below the Nyquist frequency. 
However, since the computations were very time-consuming, we limited our search mostly to the interval 
$[0;50]$\,\cd, and explored the second part of the interval only every 50 newly detected frequencies. 
For each detected frequency, the amplitude and phase were calculated by a least-squares sine fit. 
The data were then prewhitened and a new analysis was performed on the residuals. 
We did not use the option {\tt improve all}, available in the Period04 package, which computes 
a multi-sine fit.
From experience, it is known that the frequency differences derived from an individual least-squares sine fit 
relative to a multi-sine fit, depend largely on the quality of the spectral window, which in our case of CoRoT 
is characterized by a very high duty cycle.
Tests have confirmed that differences smaller than 0.001 \cd\ are typical in our case of frequencies higher  
than 0.05\,\cd, except in cases of a small-amplitude frequency within a $1/\Delta T$-bin close to a large 
amplitude one.
Our analysis was conducted until a S/N=3.9 was reached, slightly below the usual S/N=4 value 
\citep{betal93,ketal97} and corresponding to an amplitude of about 0.1\,mmag, which resulted in a detection 
of 559 frequencies.

\begin{figure}
\centering
\includegraphics[width=8cm]{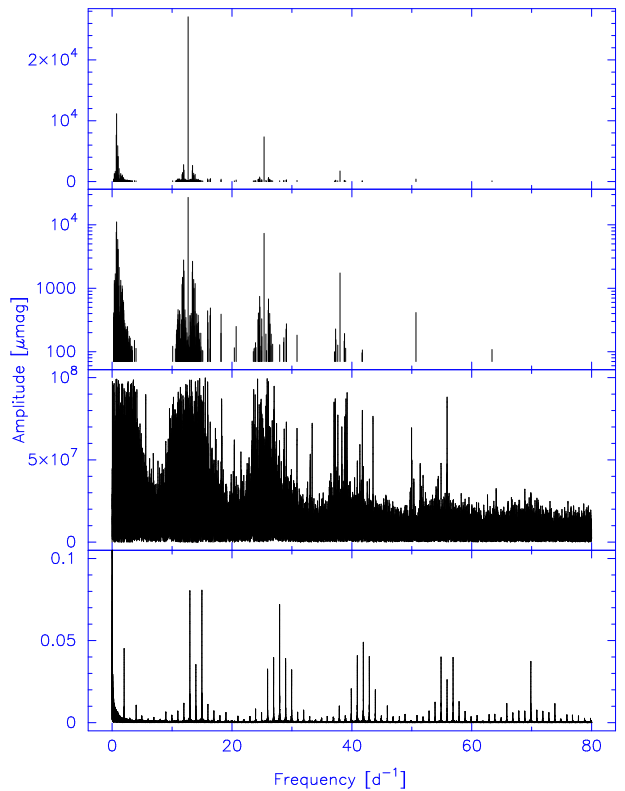}
\caption[]{Amplitude spectra computed with Period04 for the 444 retained frequencies. 
From top to bottom: the amplitudes in 
$\mu$mag, and the same with a logarithmic scale, 
the Fourier spectrum computed from the residuals,
and the spectral window normalized to unity, but plotted only to 0.1 in amplitude. 
}
\label{fig02}
\end{figure}

The uncertainties in the frequencies, amplitudes, and phases were computed with the formulae proposed by 
\cite{mo99} providing a $3\sigma$ estimate. 
However, as we used a step size of $1/20\Delta T$ in the Fourier analysis, and since the accuracies of the 
frequencies were not improved by a multi-sine fit, the uncertainties in the frequencies are certainly no smaller than 
$1/40\Delta T$. 
Therefore, the frequency uncertainties are $1/40\Delta T$ when the aforementioned formal 
$3\sigma$ estimate is smaller.
Some of the lowest frequencies are actually linked to instrumental fluctuations, which were not corrected 
by the reduction pipeline.
Once the 559 frequencies are removed, the periodogram of the residuals shows that some signal is still present 
(Fig.\,\ref{fig02}) but mainly located in high-density peak regions, and many of the peaks are actually residuals
of previously detected frequencies.

In a next step, the frequency analysis was repeated by applying SigSpec to the data filtered with a cubic spline 
function to get rid of the main CCD variations. 
However, this procedure also eliminated possible low-frequency pulsations. 
To be consistent with the Period04 results, the program was stopped after detecting 600 frequencies. 
The threshold magnitude turned out to be about 0.08\,mmag.

The results of the two methods agree fairly well. 
With SigSpec, we detected fewer low-frequencies since the data were prewhitenened with a cubic spline, but some 
additional frequencies emerged with marginal amplitudes between 0.08 and 0.1\,mmag. 
The main difference concerns the frequency values derived by the two techniques, which can amount sometimes 
to 0.004\,\cd, but remains well within the Rayleigh criterion of 0.007\,\cd. 
As we needed consistency to detect couplings, we used Period04 in the following.

The next step consisted in the sorting of stellar and non-stellar frequencies. 
As usual for CoRoT data, we eliminated both the frequencies corresponding to the orbital frequency 
$f_{\rm orb}=13.97$\,\cd\ and its harmonics, and the $2f_{\rm sid}=2.0054$\,\cd\ value due to the passage over 
the South Atlantic Anomaly, which occurs twice each sidereal day \citep{abb09}.  
Small-amplitude frequencies separated by $1/\Delta T$ from a large-amplitude frequency were ignored in the 
following as the effects of the spectral window (data set length).

As usual in the CoRoT data, variations in the properties of the CCD perturb the lowest
part of the periodogram, so we decided to analyze only frequencies
higher than 0.25\,\cd. 
We made an exception of the low amplitude frequency
$F_{144}=0.2482$\,\cd\ as it corresponds  to the coupling $f_1-f_4$. 
We tested our 0.25\cd\ limit in Sect.\,5 and found that it is realistic.

As a result, 444 frequencies were retained that have no clear instrumental origin (in Table\,1,
only the first 25 frequencies are given here and the full table is only available electronically, and 
in Fig\,\ref{fig02}). 
These frequencies can be divided into two main groups: 198 in the interval $[0.2483;3.9936]$\,\cd, corresponding 
to the typical \gd\ frequency domain, and 246 in the interval $[10.1038;63.3800]$\,\cd, 
which is typical of the \ds\ domain.
No frequency with an amplitude larger than 0.1\,mmag was detected in the interval $[4;10]$\,\cd, except 
perhaps an isolated frequency $f=5.6038$\,\cd\ with an amplitude of 0.09\,mmag. 
%table 1. List of the 444 frequencies retained (electronic form)
\begin{table*}
\begin{center}
\caption[]{The first 25 stellar frequencies (the complete table is available 
electronically).
The successive columns respectively correspond to the frequency detection order,
its value and related $S/N$, amplitude, and phase. 
We also indicate the new identification, the eventual linear combination, and finally
the eventual $n$ radial order (relative to $f_1$) of the asymptotic $g$-modes.
}
\begin{tabular}{lrrrrccr}  
\hline
\multicolumn{2}{c}{Frequency} &
\multicolumn{1}{c}{S/N} &
\multicolumn{1}{c}{A} &
\multicolumn{1}{c}{$\Phi$} &
\multicolumn{1}{c}{Ident.} &
\multicolumn{1}{c}{Linear} &
\multicolumn{1}{c}{$n$} \\
\multicolumn{1}{c}{} &
\multicolumn{1}{c}{\cd} &
\multicolumn{1}{c}{ } &
\multicolumn{1}{c}{mmag} &
\multicolumn{1}{c}{rad} &
\multicolumn{1}{c}{} &
\multicolumn{1}{c}{comb.} &
\multicolumn{1}{c}{} \\
\hline
$F_1$    & 12.6759 & 1100.7 & 27.077 & 2.259 & $F$      & -- & --  \\
$F_3$    &  0.7428 &  453.1 & 11.146 & 0.564 & $f_1$    & -- &  0  \\
$F_4$    &  0.6952 &  310.7 &  7.642 & 1.752 & $f_2$    & -- &  3  \\
$F_5$    & 25.3521 &  298.7 &  7.349 & 0.277 & $2F$     & -- & --  \\
$F_6$    &  0.9345 &  239.7 &  5.896 & 3.932 & $f_3$    & -- & -9  \\
$F_7$    &  0.9907 &  167.8 &  4.127 & 3.319 & $f_4$    & -- &-11  \\
$F_8$    &  0.9083 &  167.4 &  4.118 & 1.604 & $f_5$    & -- & -8  \\
$F_9$    &  0.9624 &  165.4 &  4.068 & 3.683 & $f_6$    & -- &-10  \\
$F_{11}$ &  0.7979 &  162.1 &  3.988 & 3.713 & $f_7$    & -- & -3  \\
$F_{12}$ &  0.8390 &  145.0 &  3.566 & 6.123 & $f_8$    & -- & -5  \\
$F_{13}$ & 11.9335 &  113.5 &  2.792 & 5.642 & $F-f_1$  & -- & --  \\
$F_{14}$ & 13.4186 &  108.3 &  2.663 & 4.104 & $F+f_1$  & -- & --  \\
$F_{15}$ &  1.1656 &   86.8 &  2.136 & 1.735 & $f_9$   & -- & --  \\
$F_{16}$ & 11.9807 &   76.2 &  1.875 & 4.814 & $F-f_2$  & -- & --  \\
$F_{17}$ & 13.3714 &   73.4 &  1.805 & 3.970 & $F+f_2$  & -- & --  \\
$F_{18}$ & 38.0279 &   70.8 &  1.742 & 4.804 & $3F$     & -- & --  \\
$F_{19}$ &  0.5279 &   69.0 &  1.697 & 0.657 & $f_{10}$ & -- & --  \\
$F_{20}$ &  0.7262 &   65.2 &  1.605 & 3.151 & $f_{11}$ & -- &  1  \\
$F_{22}$ & 11.7417 &   60.6 &  1.491 & 2.099 & $F-f_3$  & -- & --  \\
$F_{23}$ &  0.8835 &   60.3 &  1.483 & 6.217 & $f_{12}$ & -- & -7  \\
$F_{25}$ &  0.6655 &   59.1 &  1.454 & 0.456 & $f_{13}$ & -- &  5  \\
$F_{26}$ & 13.6104 &   56.0 &  1.378 & 1.123 & $F+f_3$ & -- & --  \\
$F_{28}$ &  1.4380 &   54.3 &  1.335 & 3.424 & $f_{14}$ & $f_1+f_2$ & -- \\
$F_{29}$ &  0.6786 &   54.3 &  1.335 & 2.761 & $f_{15}$ & -- & --  \\
$F_{30}$ &  0.3659 &   54.2 &  1.333 & 4.255 & $f_{16}$ & -- & --  \\

\hline
\end{tabular}
\end{center}
\label{tab1}
\end{table*}

%figure 2

This clear distinction between low- and high-frequency regions may be the consequence of a quite low angular 
rotation velocity $\Omega$. 
Following the rotation perturbation theory, the observed frequencies are indeed shifted from the co-moving 
frame ones by a quantity that is, to first order, proportional to $\Omega$. 
Hence, in the case of rapid rotators, the ``observed'' frequency spectrum might differ significantly from 
those measured in a ``co-rotating'' frame and the two frequency regions might even overlap.
It is thus imperative to obtain a spectrum of this star, which allows us to measure $v\,{\rm sin}i$ and estimate 
the rotation velocity $\Omega$. 
Unfortunately, no spectroscopy is yet available for CoRoT\,ID\,105733033. 
However, the clear gap between rather compact low- and high-frequency regions strongly argues in favor of a slow 
angular rotation, which we therefore predict. 
A low $\Omega$ would also be consistent with a rotation period of 20.57\,d, which is proposed by us in Sect.\,\ref{sec:rotation}. 
In the following, we therefore assume that the observed frequencies do not differ significantly from the co-rotating ones.

\section{The \gd\ domain}

\begin{figure}
\centering
\includegraphics[width=8.5cm]{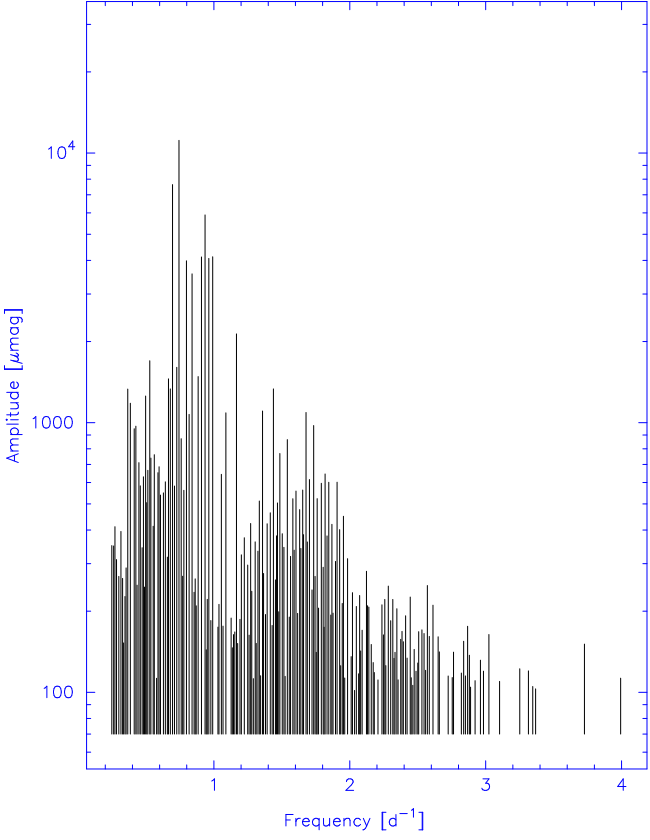}
\caption[]{Enlargement of Fig.\,2 [second] showing the detected frequencies in the \gd\ domain. 
The ordinates are on a logarithmic scale.
}
\label{fig03}
\end{figure}

\begin{figure}
\centering
\includegraphics[width=6.5cm,angle=-90.]{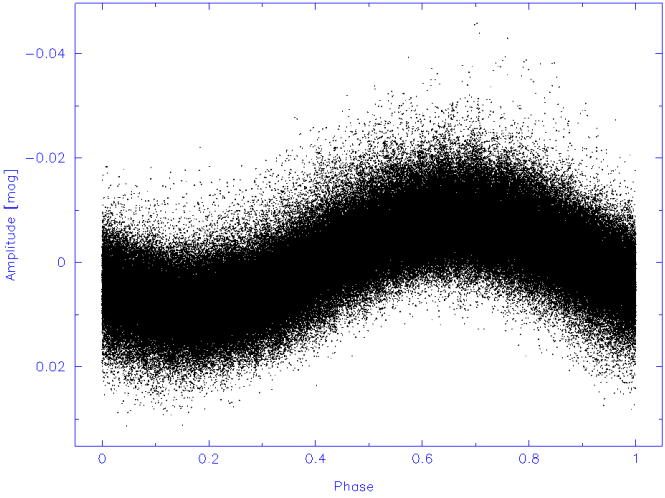}
\caption[]{Data phased with the main \gd\ frequency $f_1=0.7428$\,\cd.
}
\label{fig04}
\end{figure}

From the Fourier analysis, we found 198 frequencies in the range $[0.2483;3.9936]$\,\cd, of which 176 were detected 
between 0.25 and 2.6\,\cd\ (Fig.\,\ref{fig03}). 
In the following, we re-numbered the 198 frequencies in order of decreasing amplitude, starting from $f_1$.
%figure 3
The main frequency is $f_1=0.7428$\,\cd\, with an amplitude $A_1=111.46$\,mmag. 
Data phased with this frequency are represented in Fig.\,\ref{fig04}.
%figure 4
The shape of the curve is nearly sinusoidal, hence only 2$f_1$ is detected with an amplitude of 
0.77\,mmag.

We identified 17 cases in which a linear combination of detected \gd\ frequencies coincide to within 
$\Delta =0.001$\,\cd\ with 17 other detected frequency (Table\,2). 
This $\Delta$ refers to the uncertainty in the lowest amplitude frequencies (see Table\,1). 

Another interesting finding is the presence of coupling between $p$- and $g$-modes (see Sect.\,5), 
where we found for 82\,\% of the 50 candidates with the largest amplitudes a frequency difference between 
coupled modes and observed frequencies of less than 0.001\,\cd. 
On the other hand, in the case of the 100 lowest amplitude $g$-modes (for which no coupling is expected) 
we found in only 1.5\,\% of the possibilities a difference smaller than 0.001\,\cd. 
We interpret this finding as support of the chosen value of $\Delta$=0.001\,\cd.
A few expected combinations were not detected, e.g. $f_1+f_5$, as they were masked by pulsation frequencies 
with higher amplitudes (within a distance smaller than $1/\Delta T$). 
After elimination of harmonics and coupled frequencies, 180 independent frequencies of \gd\ type were 
retained.

%table 2. CL of gd type.
\begin{table}
\begin{center}
\caption[]{Linear combinations of frequencies detected in the \gd\ domain. 
Columns are respectively the label of the frequency from Table\,1 together with its new label, 
the linear combination, and the differences between the linear combination and the actually measured 
values for Period04 and SigSpec [$\times 10^{-5}$\,\cd].}
\begin{tabular}{llcrr}  
\hline
\multicolumn{2}{c}{Frequency} &
\multicolumn{1}{c}{Combination} &
\multicolumn{1}{c}{$\Delta$ Period04} &
\multicolumn{1}{c}{$\Delta$ SigSpec} \\
\multicolumn{2}{c}{} &
\multicolumn{1}{c}{} &
\multicolumn{1}{c}{[$10^{-5}$\,\cd]} &
\multicolumn{1}{c}{[$10^{-5}$\,\cd]} \\
\hline
$F_{28}$  & $f_{14}$  & $f_1 +f_2$    & 4   &  13 \\
$F_{40}$  & $f_{20}$  & $f_1 +f_3$    & 39  &  372 \\
$F_{47}$  & $f_{23}$  & $f_1 +f_4$    & 73  &  -26 \\
$F_{55}$  & $f_{27}$  & $f_1 +f_7$    & -31 &  30 \\
$F_{89}$  & $f_{51}$  & $f_1 +f_8$    & 66  & not detected \\
$F_{110}$ & $f_{58}$  & $f_1 +f_{10}$ & -35 & 88 \\
$F_{256}$ & $f_{112}$ & $f_1 +2f_3$   & 70  & 123 \\
$F_{57}$  & $f_{28}$  & $2f_1$        & -31 &  157 \\
$F_{470}$ & $f_{175}$ & $2f_1+f_2$    & -64 &  19 \\
$F_{358}$ & $f_{149}$ & $2f_1+f_5$    & 4   &  194 \\ 
$F_{218}$ & $f_{97}$  & $2f_1+f_7$    & 5   &  -194 \\
$F_{346}$ & $f_{146}$ & $2f_1+f_9$    & 66  &  127 \\
$F_{83}$  & $f_{47}$  & $f_2 +f_5$    & 73  &  17 \\
$F_{286}$ & $f_{123}$ & $f_2 +f_9$    & 33  &  377 \\
$F_{134}$ & $f_{68}$  & $f_2 +f_{10}$ & 68  &  -48 \\
$F_{412}$ & $f_{165}$ & $2f_2+f_3$    & 75  &  -36 \\
$F_{122}$ & $f_{62}$  & $f_3 +f_4$    & 2   &  68 \\
$F_{144}$ & $f_{71}$  & $f_4 -f_1$    & 35  &  343 \\
\hline
\end{tabular}
\end{center}
\label{tab2}
\end{table}

\begin{figure}
\centering
\includegraphics[width=8cm]{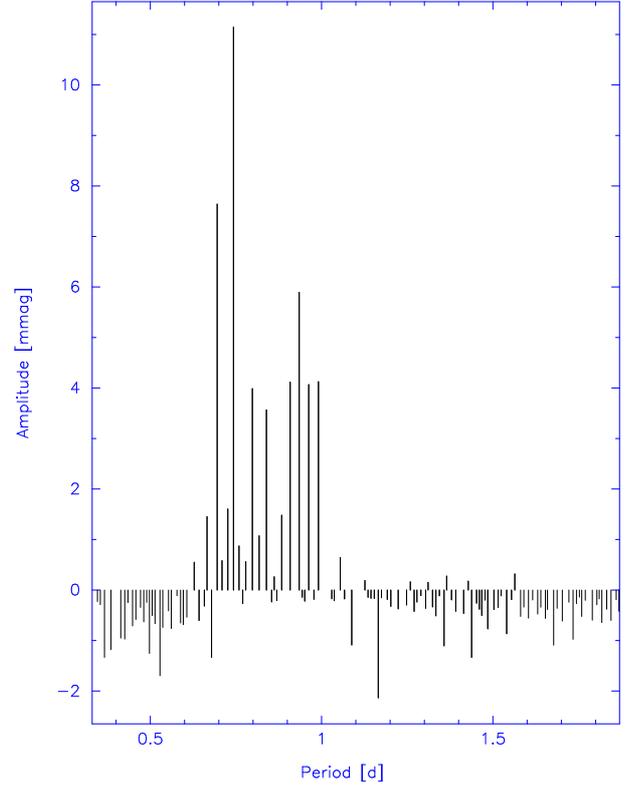}
\caption[]{Plot of the frequencies detected in the interval [0.4;1.8]\,\cd.
"Positive" amplitude frequencies represent the identified asymptotic frequency series,
while the "negative" ones are the other detected frequencies in this interval.
}
\label{fig04bis}
\end{figure}

\begin{figure}
\centering
\includegraphics[width=8cm]{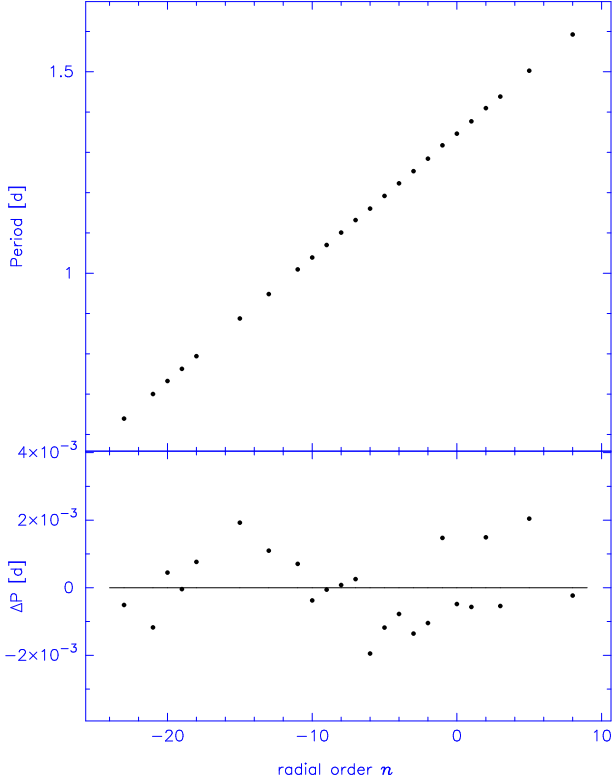}
\caption[]{\gd\ periods that obey the asymptotic period spacing law as a function of radial order $n$ with 
an arbitrary zero point [top]. 
The bottom part shows the residuals of this period spacing; a straight line has been added for better visibility.
}
\label{fig05}
\end{figure}

In the case of large radial numbers $n$, the first order asymptotic theory \citep{t80} predicts for 
$g$-modes of the same degree $\ell$, a constant period spacing between consecutive radial orders $n$. 
Searching in Table\,1 for equidistant \gd\ periods, we found a series of 24 asymptotic ones with a mean 
separation of $\Delta P=0.03074\pm 0.00005$\,d (Fig.\,\ref{fig04bis}).
%figure 5
The periods extend from 0.693\,d to 1.593\,d with the highest concentration being between 1.009\,d and 1.439\,d. 
In this region, we detected 15 consecutive and equidistant periods among the 18 possible. 
Most of them have relatively large amplitudes, and are indeed among the largest amplitude g-mode 
frequencies detected. 
Further away from the central region, the amplitudes decrease, but we detected 9 other periods 
fitting the sequence. 
Several values are missing in the asymptotic series probably owing to their low amplitudes. 
The 24 retained periods correspond to $n$-values ranging from $-23$ to $+8$, with the largest amplitude 
period $P_1 = 1.3470$\,d ($f_1=0.7428$\,\cd) being arbitrarily taken as order zero. The sequence is 
listed in Table\,3 and illustrated in Fig.\,\ref{fig05}.

%figure 5
%table 3. Period spacings in GDor frequencies
\begin{table}
\begin{center}
\caption[]{List of the asymptotic frequencies. Columns are respectively the label of the frequency from 
Table\,1 together with its \gd\ label, the corresponding period value [d], the related amplitude [mmag], 
and the order $n$ of the pulsation mode (arbitrarily shifted to $n=0$ for the main frequency $f_1$).}
\begin{tabular}{llrrr}  
\hline
\multicolumn{2}{c}{Frequency} &
\multicolumn{1}{c}{Period} &
\multicolumn{1}{c}{Amplitude} &
\multicolumn{1}{c}{$n$} \\
\hline
$F_{163}$ & f$_{79 }$ &   0.6393 & 0.32 & -23\\
$F_{313}$ & f$_{130}$ &   0.7001 & 0.18 & -21\\
$F_{191}$ & f$_{88 }$ &   0.7325 & 0.28 & -20\\
$F_{365}$ & f$_{150}$ &   0.7627 & 0.15 & -19\\
$F_{339}$ & f$_{144}$ &   0.7943 & 0.16 & -18\\
$F_{296}$ & f$_{126}$ &   0.8877 & 0.19 & -15\\
$F_{70 }$ & f$_{36 }$ &   0.9483 & 0.64 & -13\\
$F_{7  }$ & f$_{4  }$ &   1.0094 & 4.13 & -11\\
$F_{9  }$ & f$_{6  }$ &   1.0391 & 1.07 & -10\\
$F_{6  }$ & f$_{3  }$ &   1.0701 & 5.90 &  -9\\
$F_{8  }$ & f$_{5  }$ &   1.1010 & 4.12 &  -8\\
$F_{23 }$ & f$_{12 }$ &   1.1319 & 1.48 &  -7\\
$F_{207}$ & f$_{93 }$ &   1.1604 & 0.26 &  -6\\
$F_{12 }$ & f$_{8  }$ &   1.1919 & 3.57 &  -5\\
$F_{42 }$ & f$_{22 }$ &   1.2231 & 1.07 &  -4\\
$F_{11 }$ & f$_{7  }$ &   1.2532 & 3.99 &  -3\\
$F_{82 }$ & f$_{46 }$ &   1.2843 & 0.56 &  -2\\
$F_{54 }$ & f$_{26 }$ &   1.3176 & 0.87 &  -1\\
$F_{3  }$ & f$_{1  }$ &   1.3463 &11.15 &   0\\
$F_{20 }$ & f$_{11 }$ &   1.3770 & 1.61 &   1\\
$F_{80 }$ & f$_{44 }$ &   1.4098 & 0.58 &   2\\
$F_{4  }$ & f$_{2  }$ &   1.4385 & 7.64 &   3\\
$F_{25 }$ & f$_{13 }$ &   1.5026 & 1.45 &   5\\
$F_{86 }$ & f$_{48 }$ &   1.5925 & 0.55 &   8\\
\hline
\end{tabular}
\end{center}
\label{tab3}
\end{table}

\section{The \ds\ domain}

\begin{figure}
\centering
\includegraphics[width=6.5cm,angle=-90.]{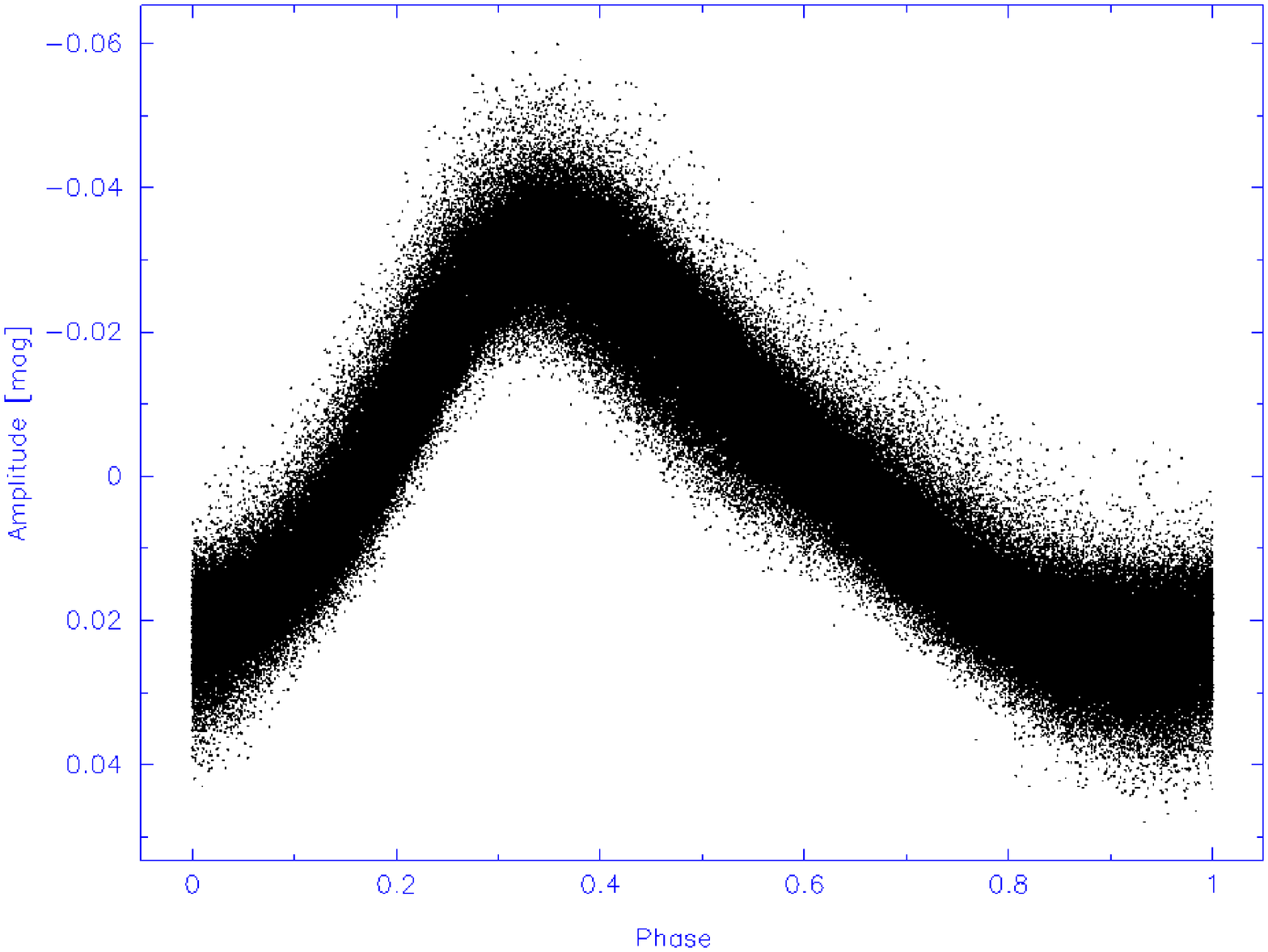}
\caption[]{Data phased with the main \ds\ frequency $F=12.6759$\,\cd.
}
\label{fig07}
\end{figure}

In the \ds\ frequency domain, we retained 246 frequencies in the interval $[10.1038;63.3800]$\,\cd\ 
(Table\,1, Fig.\,\ref{fig07}). 
One frequency clearly dominates the spectrum, $F=12.6759$\,\cd, with an amplitude $A_1=27.08$\,mmag.
Figure\,\ref{fig07} represents a light curve that has been phased with frequency $F$. 
The curve is very asymmetric and unsurprisingly harmonics up to 5$F$ were detected. 
The Fourier parameters $\phi_{21}=4.03$\,rad (phase difference of first overtone and twice the phase of 
fundamental frequency) and $R_{21}=0.26$ (the amplitude ratio of first overtone to fundamental frequency) 
were in perfect agreement with those of fundamental radial pulsators \citep{p01}. 
Hence, we can conclude that $F$ is a radial mode.
%figure 7

To corroborate the identification of this pulsation mode, we have only the measurements provided by the 
EXODAT database \citep{dmm09}, which are indicative of an A5V spectral type. 
On the basis of this classification, the star would have an absolute magnitude of $M_V=+2.0$\,mag and an effective 
temperature of $\teff = 8000$\,K. 
The period-luminosity relation given by \citet{tbd02} supports the identification of the main frequency 
$F$ as a radial fundamental mode. 
Likewise, using the equation given by \citet{pch08} $M_V= -1.83 (\pm 0.08)-3.65( \pm 0.07) \log P$, 
we also obtain for the absolute magnitude: $M_V= +2.20 (\pm 0.16)$\,mag. 
%figure 8
\begin{figure}
\centering
\includegraphics[width=8.5cm]{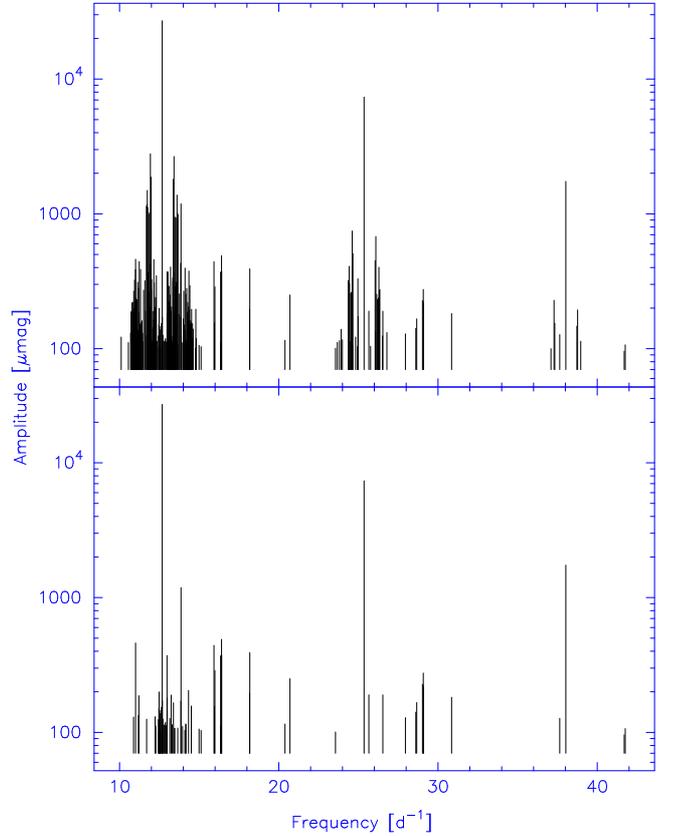}
\caption[]{Top: Enlargement of Fig.\,2 [second] showing the detected frequencies in the \ds\ domain. 
Bottom: the \ds\ frequencies without linear combinations with \gd\ frequencies. 
The ordinates are on a logarithmic scale.
}
\label{fig08}
\end{figure}

The most remarkable property of CoRoT\,ID\,105733033 is the coupling of most of the \ds\ frequencies with 
\gd\ ones: 
the first tens of \gd\ frequencies $f_i$ are present in the \ds\ domain as $F \pm f_i$. 
We also detected $2F \pm f_i$ and $3F \pm f_i$ for the \gd\ frequencies with the largest amplitudes. 
For the 40 first \gd\ frequencies $f_i$, both $F - f_i$ and $F + f_i$ frequencies are present 
except for a few ambiguous cases.
Coupling with lower amplitude \gd\ frequencies (up to $f_{100}$) were also detected as $F-f_i$ or $F+f_i$, 
but not so regularly. 
We were able to detect 146 couplings within the limit of $\Delta f = 0.001$\,\cd: 110 for $F \pm f_i$, 
30 for $2F \pm f_i$, and 6 for $3F \pm f_i$. 
The three first groups of coupling are presented in Table\,4, the complete list being presented in Table\,1.
%table 4. List of the coupled \gd\ and \ds\ frequencies
\begin{table}
\begin{center}
\caption[]{List of the first three groups of coupled \gd\ and \ds\ frequencies. 
Columns are respectively the frequency label of the first column of Table\,1, the related coupling, 
$\Delta=|F \pm f-F_{\rm measured}|$ in units of $10^{-5}$\,\cd, and the amplitude ratio 
$A_{f_i}/(A_{kF \pm f_i})$. 
We recall that the Rayleigh frequency resolution (1/$\Delta$ T) corresponds to $70\cdot 10^{-5}$\,\cd. }
\begin{tabular}{lcrr}  
\hline
\multicolumn{1}{c}{Frequency} &
\multicolumn{1}{c}{Coupling} &
\multicolumn{1}{c}{$\Delta$} &
\multicolumn{1}{c}{Ratio} \\
\hline
$F_{13}$  & $F-f_1 $ & 34.5 &   4.00 \\
$F_{14}$  & $F+f_1 $ &  0.3 &   4.19 \\
$F_{16}$  & $F-f_2 $ &  0.4 &   4.09 \\
$F_{17}$  & $F+f_2 $ & 38.9 &   4.22 \\
$F_{22}$  & $F-f_3 $ & 35.1 &   3.96 \\
$F_{26}$  & $F+f_3 $ &  0.6 &   4.28 \\
\hline
$F_{59}$  & $2F-f_1$ & 11.2 &  14.93 \\
$F_{64}$  & $2F+f_1$ & 34.2 &  16.40 \\
$F_{94}$  & $2F-f_2$ &  4.0 &  15.08 \\
$F_{104}$ & $2F+f_2$ & 40.4 &  16.97 \\
$F_{119}$ & $2F-f_3$ &  0.1 &  14.46 \\
$F_{121}$ & $2F+f_3$ & 34.9 &  14.69 \\
\hline
$F_{233}$ & $3F-f_1$ & 18.1 &  48.73 \\
$F_{284}$ & $3F+f_1$ &  0.0 &  57.52 \\
$F_{355}$ & $3F-f_2$ & 34.9 &  49.32 \\
$F_{378}$ & $3F+f_2$ & 68.6 &  52.04 \\
$F_{555}$ & $3F-f_3$ &  0.1 &  58.95 \\
$F_{490}$ & $3F+f_3$ & 43.8 &  51.98 \\
\hline
\end{tabular}
\end{center}
\label{tab4}
\end{table}

The amplitude ratios of the \gd\ frequencies to the associated high frequencies of \ds\ 
characteristics are remarkably constant, being $4.1\pm 0.1$ ($A_{f_i}/(A_{F \pm f_i})$), 
$15\pm 1$ ($A_{f_i}/(A_{2F \pm f_i}))$, and $53\pm 4$ ($A_{f_i}/(A_{3F \pm f_i})$). 
The amplitudes of the $F - f_i$ frequencies were slightly larger (1.3\,\%) than those of $F + f_i$.
We also note that the beatings of $F \pm f_i$ frequencies are in phase with the $f_i$ frequencies i.e.,
the highest amplitude of the beating of $F$ and $F \pm f_i$ occurs when the light curve of $f_i$ reaches a maximum.

Thus, it appears that a fundamental property of this star is the coupling between $g$- and $p$-modes. 
This coupling was previously predicted for the Sun \citep{kjh93}. 
Following this study, the low-frequency $g$-modes, trapped in the stellar interior, induce slow periodic 
thermodynamic perturbations in the envelope, leading to a frequency modulation of the outer $p$-modes. 
This causes the formation of a pair of spectral sidelobes (not to be confused with sidelobes caused by the 
spectral window!) that are symmetric relative to the unperturbed $p$-mode frequency. 
This is exactly what we see for our $F \pm f_i$ peaks. 
A complementary study \citep{l01} shows that, in principle, a spectrum of sidelobes would appear to be 
centered on each $p$-mode frequency and that the spectral structure may be complicated by the presence 
of several $g$-modes. 
The predicted amplitude ratio of the first sidelobe $A_s$ to the central $p$-mode $A_p$ is 
$\frac{A_s}{A_p}=\frac{1}{2}\frac{A_{F} \pm A_{f_i}}{A_{f_i}} \approx 0.125$, 
but this ratio was not measured by us. 

On the other hand, measured sidelobe amplitudes are directly linked to the amplitude of the respective 
``parent'' $g$- and $p$-modes. 
The first non-radial mode of \ds-type ($p_1$) also contains sidelobes. 
The amplitude ratio of $F$ to $p_1$ is 23, where the sidelobes of the $p_1\pm f_i$ peaks are ``only'' nine
times smaller than those corresponding to $F\pm f_i$. 
Hence, the relationship does not seem to be linear.
In conclusion, except for the influence of the amplitudes of the concerned modes, the coupling mechanism 
proposed by \citet{kjh93} and \citet{l01} operates in this star. 
Initiated first as a help to the detection of $g$-modes in the Sun, there is no obvious reason 
for not assuming that this mechanism operates in other stars.

Without all the coupling frequencies 100, \ds-type frequencies remained. 
For a list of\ ``pure'' \ds\ frequencies, we also discarded frequencies 
with $|kF \pm f_i - F_{\rm measured}| \leq \Delta=0.0015$\,\cd (instead of our standard 
$\Delta=0.0010$\,\cd), if their amplitude ratio was compatible with the value of three obtained
in our ``standard'' case. 
We derived 25 of these possible couplings, which are all of $F \pm f_i$ type. 

We used this interaction mechanism in order to test the low frequency
region.
If \gd-type frequencies, $f_i$, lower than 0.25\,\cd\ and with amplitudes larger than 
0.4\,mmag were to exist, we should detect them as $F \pm f_i$.
No such frequencies were detected, so we conclude that the lower limit for \gd\ frequencies really is 
about 0.25\,\cd.

The physical characteristics of the star (mass, effective temperature, radius, 
evolutionary state\ldots), are uncertain, as was previously mentioned at the beginning of this section. 
Using the values computed by \citet{f81}, the frequencies associated with the first overtones 1O, 2O, and 
3O should be close to respectively 16.55, 20.66, and 24.85\,\cd. 
Values associated with the 1O and 2O modes are located in regions that have a low density of frequencies, 
and the frequencies $F_{98}=16.4019$\,\cd ($A_{98}=0.49$\,mmag) and 
$F_{215}=20.7007$\,\cd ($A_{215}=0.25$\,mmag) could correspond to these modes.
As neither the fundamental stellar parameters, nor the true errors of the models, are known with sufficient accuracy, 
we can only speculate that the overtones are a correct interpretation.

The non-radial \ds\ frequencies were renamed in order of decreasing amplitude starting from 
$p_1 \equiv F_{33}$ (see also Table\,1).
They are presented in the lower part of Fig\,\ref{fig08}. 
These frequencies are dominated by $p_1=13.8566$\,\cd\ with an amplitude of 1.18\,mmag 
and all the others have amplitudes smaller than 0.5\,mmag. 
We detected three couplings between $p_1$ and the \gd\ frequencies $f_3$, $f_5$, and $f_6$ 
according to $p_1 + f_i$. 
The mean amplitude ratio $A_{p_1} /(A_{p_1+f_i})$ is 35 i.e., 
about nine times larger than the corresponding ratio associated to $F$. 

We also detected 9 couplings between $F$ and other \ds\ type frequencies. 
Therefore, we end up with 59 independent \ds\ frequencies, three of which are probably associated with radial 
modes, and the 56 others with non-radial modes.

\section{Rotation}    \label{sec:rotation}

We searched for equidistant frequencies that could be related to rotational splitting. 
The only possibilities we found were quadruplet components formed by $p_{31} \equiv F_{381}=12.5776$\,\cd, 
$p_{29} \equiv F_{360}=12.6276$\,\cd, $F =12.6759$\,\cd\, and $p_{41} \equiv F_{444}=12.7231$\,\cd. 
The separations are respectively 0.0500, 0.0483, and 0.0473\,\cd. 
The radial fundamental mode $F$ cannot of course lead to any rotational splitting, but its high 
amplitude may conceal a non-radial $\ell=2$ mode that could be responsible for the aforementioned detected 
quadruplet.
Another interpretation is that the star belongs to a binary system, and that the equidistant peaks are actually orbital 
sidelobes to the central pulsation frequency modulated by the Doppler effect, as proposed by \citet{sk12}.

Among the frequencies below 0.25\,\cd\ (see Sec.\,3) was a frequency at 
0.04862\,\cd\ that has an amplitude of $A=1.46$\,mmag, which is close to the separation values 
mentioned in the previous paragraph. 

\begin{figure}
\centering
\includegraphics[width=6.5cm]{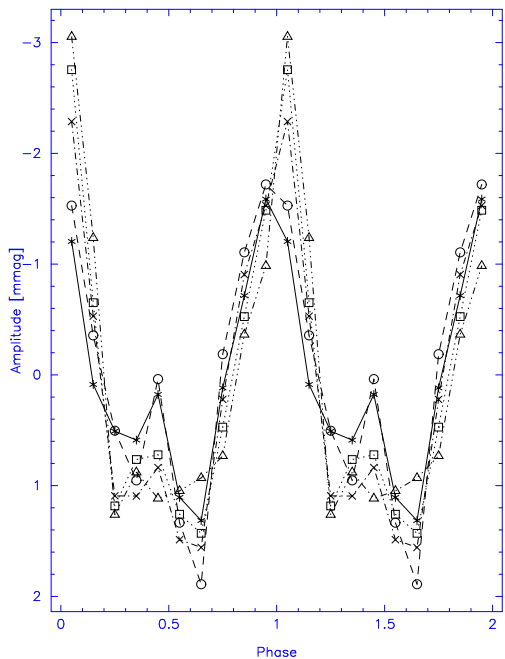}
\caption[]{Binned data phased with the candidate rotation frequency $f=0.0486$\,\cd,
corresponding to a rotation period of 20.57\,d.
Different symbols refer to different rotation cycles and data averaged in steps of 0.1 in phase. 
}
\label{fig09}
\end{figure}

A possible interpretation of this low frequency could be the rotation of the star with a period of 20.57\,d 
causing light variations associated for instance with surface inhomogeneities.
We note that a phase plot (Fig.\,\ref{fig09}) of the residuals after prewhitening with the pulsation frequencies, 
when using the candidate rotation period looks reasonable.
This rotation period would also be long enough to account for the two clearly distinct groups of \gd\ and 
\ds\ frequencies because the observed frequency spectrum should not differ significantly from the co-rotating one.
Our speculations concerning a rotation period of course have to be taken with the utmost caution
as the very low frequency range is populated by signal caused by instrumental effects, imperfect data reduction, 
$g$-modes, and combination frequencies. 
A dedicated spectroscopic observing program focused on the detection of rotation and eventual binarity effects 
in this star would be needed to resolve the ambiguities. 
However, the evidence presented in this paper is both consistent and plausible.
%figure 9

\section{Summary}

The star CoRoT\,ID\,105733033 appears to be a very powerful test case for hybrid star studies. 
First, its frequency spectrum is quite rich.
Second, two separate domains are clearly detected, with 198 frequencies in the range of 
$[0.2483;3.9936]$\,\cd, which is typical for $g$-modes, and 246 frequencies in the range 
of $[10.1038;63.3800]$\,\cd, typical of $p$-modes.
Among these frequencies, 180 and 59 are respectively neither combination frequencies 
nor caused by coupling between $g$ and $p$-modes, hence seem to represent pure \gd\ and 
\ds\ pulsation modes.
The sole exception is the frequency $f=5.6038$\,\cd, but with an amplitude slightly smaller than 
0.1\,mmag, which is our chosen lower magnitude limit. 
This gap between the two domains is a strong indication of slow rotation, since the modes 
in the observer's frame are not mixed.

We found 17 linear combinations among the \gd\ frequencies. 
More are probably present, but the time coverage of our data may not be sufficiently long to detect them. 
The asymptotic theory is very well-illustrated by this star, which has 24 periods that are 
equidistant at intervals of $\Delta P=0.03074$\,d.

The high-frequency region shows a dominant peak $F$ that could be identified with the radial 
fundamental mode.
It is most remarkable that of the 246 \ds\ frequencies, only about 70 ($\approx 30$\,\%) 
seem really independent of a \gd\ frequency. 
Twice as many (146) couplings of the form $kF\pm f_i$ were detected. 
Depending on $\Delta$, which is the minimum frequency difference to allow for a clear determination of 
neighbouring frequencies, this number could increase to 171 and more. 
Additional coupled frequencies are certainly present, but hidden in the noise or too close to a 
frequency of larger amplitude.

\citet{km10} claimed that the several hundreds of \ds-type frequencies detected in A-F type 
stars observed by CoRoT and Kepler could not all be attributed to pulsation. 
Since the observable amplitudes of modes with $\ell$ larger than 4, which are required to explain the large 
number of pulsation modes, suffer strong cancellation effects when integrated over the stellar 
disk, only some tens of frequencies should be detected. 
The authors demonstrated that many low amplitude frequencies could alternatively be the signature of non-white 
granulation background noise and that fewer than about 100 of the frequencies of the investigated stars 
are actually stellar $p$-modes. 
In our case, we detected ``only'' 59 independent modes.
%WW I am afraid I was too quick in this statement, as I remember that of the many frequencies 
%Breger has detected he also attributes some of them to linear combinations. What shall we write?? 
%Is "compatible" vague enough? But the hint to Kepler in the next sentence reduces the problem. 
%Leave it to the referee?
Therefore, only $\ell$-values up to 5 are necessary to attribute the frequencies to p-modes, but 
significantly higher $\ell$ values are discussed nowadays for Kepler data.
However, as the granulation level is not known for CoRoT\,ID\,105733033, we cannot be entirely sure 
that all the 59 \ds- type frequencies are really associated with pulsation.

We also detected equidistant frequencies very close to $F$ that could be interpreted as a signature 
of rotational splitting or Doppler effect modulation. 
If this were indeed true, the rotation or binary period of the star would be 20.57\,d.

We have clearly established the influence of the \gd\ variations on the \ds\ behaviour, and convincingly 
confirmed the thermodynamical and mechanical interactions between the two $p$- and $g$-mode cavities, 
as predicted many years ago as a tool for detecting $g$-modes in the Sun \citep{kjh93,l01}.

We now know that the instability strips of \gd\ and \ds\ stars are not as 
clearly defined as previously 
thought \citep{umg11}, and the star CoRoT\,ID\,105733033 might be the Rosetta stone for hybrid stars, 
where coupling similar to those detected in the present paper should be systematically searched 
for. 
We are presently studying other candidates in the CoRoT exofield \citep[see e.g.][]{mcb09,hrm10}.

\begin{acknowledgements}
The authors wish to thank E. Poretti and F.-X. Schmider for useful comments.
W.W. Weiss is supported by the Austrian Research Fond (project P22691-N16).
The authors thank their anonymous referee for his helpful suggestions.
\end{acknowledgements}

\end{document}